\documentclass[preprintnumbers,amsmath,amssymb,floatfix,11pt,one column,prd,superscriptaddress,nofootinbib,showpacs,showkeys]{revtex4}
\usepackage{graphicx}
\usepackage{epsfig}
\usepackage{bm}
\usepackage{amsfonts}
\usepackage{amsfonts,amssymb,amsmath}
\usepackage[hang]{subfigure}

\def\ben{\begin{equation}}
\def\een{\end{equation}}

 \def\bd{\begin{document}} \def\ed{\end{document}}
\def\ds{\documentstyle} \let\fr=\frac \let\bl=\bigl \let\br=\bigr
\let\Br=\Bigr \let\Bl=\Bigl
\let\bm=\bibitem
\let\na=\nabla
\let\pa=\partial \let\ov=\overline
\newcommand{\be}{\begin{equation}}
\newcommand{\ee}{\end{equation}}
\def\ba{\begin{array}}
\def\ea{\end{array}}
\def\ft#1#2{{\textstyle{\frac{\scriptstyle #1}{\scriptstyle #2} } }}
\def\fft#1#2{{\frac{#1}{#2}}}
\def\del{\partial}
\def\vp{\varphi}
\def\sst#1{{\scriptscriptstyle #1}}
\def\oneone{\rlap 1\mkern4mu{\rm l}}
\def\td{\tilde}
\def\wtd{\widetilde}
\def\ie{{\it i.e.\ }}
\def\dalemb#1#2{{\vbox{\hrule height .#2pt
        \hbox{\vrule width.#2pt height#1pt \kern#1pt
                \vrule width.#2pt}
        \hrule height.#2pt}}}
\def\square{\mathord{\dalemb{6.8}{7}\hbox{\hskip1pt}}}
\newcommand{\ho}[1]{$\, ^{#1}$}
\newcommand{\hoch}[1]{$\, ^{#1}$}
\newcommand{\bea}{\setlength\arraycolsep{2pt} \begin{eqnarray}}
\newcommand{\eea}{\end{eqnarray}}
\newcommand{\ra}{\rightarrow}
\newcommand{\lra}{\longrightarrow}
\newcommand{\Lra}{\Leftrightarrow}
\newcommand{\bp}{\tilde \beta^\prime}
\newcommand{\tr}{{\rm tr} }
\newcommand{\Tr}{{\rm Tr} }
\def\0{{\sst{(0)}}}
\def\1{{\sst{(1)}}}
\def\2{{\sst{(2)}}}
\def\3{{\sst{(3)}}}
\def\4{{\sst{(4)}}}
\def\5{{\sst{(5)}}}
\def\6{{\sst{(6)}}}
\def\7{{\sst{(7)}}}
\def\8{{\sst{(8)}}}
\def\m{{\sst{(m)}}}
\def\n{{\sst{(n)}}}
\def\cA{{{\cal A}}}
\def\cB{{{\cal B}}}
\def\cF{{{\cal F}}}
\def\cG{{{\cal G}}}
\def\cH{{{\cal H}}}
\def\tV{\widetilde V}
\def\tW{\widetilde W}
\def\tH{\widetilde H}
\def\tE{\widetilde E}

\def\tF{\widetilde F}
\def\tA{\widetilde A}
\def\im{{{\rm i}}}
\def\tY{{{\wtd Y}}}
\def\ep{{\epsilon}}
\def\vep{{\varepsilon}}
\def\bD{{{\bar D}}}
\def\R{{{\mathbb R}}}
\def\C{{{\mathbb C}}}
\def\H{{{\mathbb H}}}
\def\CP{{{\mathbb C}{\mathbb P}}}
\def\RP{{{\mathbb R}{\mathbb P}}}
\def\Z{{{\mathbb Z}}}
\def\bA{{{\mathbb A}}}
\def\bB{{{\mathbb B}}}
\def\bC{{{\mathbb C}}}
\def\bD{{{\mathbb D}}}
\def\bE{{{\mathbb E}}}
\def\bZ{{{\mathbb Z}}}
\def\Re{{{\frak{Re}}}}
\def\Im{{{\frak{Im}}}}
\def\cosec{{\,\hbox{cosec}\,}}
\def\Gm{{\Gamma_{\!\! -}}}
\def\Gp{{\Gamma_{\!\! +}}}
\def\stan{{standard }}
\def\nonstan{{supernumerary }}
\def\p{{\partial}}
\def\kdel#1{{\fft{\del}{\del#1}}}

\def\bog{{Bogomolny }}
\def\om{{\omega}}

\newcommand{\nnr}{\nonumber \\}
\newcommand{\pd}{\partial}
\newcommand{\ud}{\textrm{d}}
\newcommand{\dTH}{T^{\prime \, 0}_\textrm{H}}
\newcommand{\dOi}{\Omega^{\prime \, 0}_i}
\newcommand{\bx}{{\bf x}}
\begin{document}

\title{ Construction of a Holographic Superconductor in F(R) Gravity  }

\author{\textbf{ D. Momeni}}

 \affiliation{Eurasian International Center
for Theoretical Physics, Eurasian National University, Astana
010008, Kazakhstan}

\author{\textbf{M. Raza}}
\affiliation{Department of Mathematics, COMSATS Institute of Information
Technology, Sahiwal, Pakistan}

\author{\textbf{ R. Myrzakulov}}

\affiliation{Eurasian International Center
for Theoretical Physics, Eurasian National University, Astana
010008, Kazakhstan}

\begin{abstract}
We construct  a toy model for holographic superconductor with non linear Maxwell field in the frame of modified gravity. By probe the bulk background by non linear Maxwell fields we show that superconductivity  happens under a specific critical temperature. The effect of the non linear Maxwell field and non linear curvature corrections have been studied by analytical matching methods. We conclude that the non linearity in Maxwell field and curvature coupling  make condensation harder.

\end{abstract}
\pacs{ 11.25.Tq, 04.70.Bw, 74.20.-z}
 \keywords{Gauge/string duality; Theories and models of superconducting states; Modified $\mathcal{F}(R)$ gravity.}

\newpage
\maketitle
\section{Introduction}
The great discovery of Maldacena , anti de Sitter/conformal field theory (AdS/CFT) correspondence \cite{Maldacena}, finds many applications in systems with strong interactions
\cite{Herzog1}. The  AdS/CFT correspondence presents a possibility to find dual quantities  on the boundary using the classical black hole solutions of  gravitational bulk. This identification exists in the strong regime of a gauge theory (super Yang-Mills theory) with gauge coupling constant $g$ and in large numbers of colors $N_c\rightarrow\infty$. The idea is how we can identify the quantum fields (which they live on the flat boundary and described by CFT) to a weakly coupled classical system (black hole) in bulk. If the dimension of bulk be $d$, the Maldacena conjecture relates it to a $d-1$ dimensional gauge theory on boundary. Quantum fields on the boundary behave likes the plasma. To keep the system in thermal equilibrium usually we write
$$
T_{BH}=T_{CFT}
$$
Where, the first is horizon killing temperature and the second , temperature that appears in the CFT partition function via $Z_{CFT}\propto e^{-F/T_{CFT}}$. Energy momentum tensor of conformal theory has a quantum expectation value. To have a full complete description we must identify it to a finite correspondence quantity in the bulk. The energy momentum tensor of conformal boundary has zero trace and it's two point function is related to the central charge. The quantum expectation value of this energy momentum tensor needs a renormalization. Such renormalization presented from holography point of view. If the gravity part in bulk be classical general relativity (GR) action, the appropriate quantity for holographic renormalization is a counter term, proposed firstly by Gibbons-Hawking. But if we used higher order gravity corrections , this solver tool for renormalization depends to several parameters.  As an example in the applied physics any strongly correlated system with scaling invariance can be a good candidate to test this conjecture. One famous example of such systems is type-II superconductor in the condensed matter theory. The usage of the AdS/CFT conjecture in condensed matter  physics specially in strongly correlated systems and plasma is called as AdS/CMT duality.  First time, Hartnoll and his group considered solutions of the holographic superconductor with a nonzero supercurrent\cite{Hartnoll}. There was no instability in their model exactly because the effective scalar mass has a lower bound 
\emph{Breitenlohner-Freedman (BF) bound}  but free of 
instability \cite{BF}. The numerical solutions of system of coupled non linear scalar and Abelian fields, showed the existence of a supoerconductive phase in which system evolves under a critical temperature to the superconductive form. From the conformal field theory window,  the superconductivity  on the boundary is described by existence  of a charged scalar  condensate  for  temperatures $T<T_{c}$ with a definite conformal dimension.
This mechanism  inspired directly from  discovery of spontaneous symmetry breaking in the
presence of horizon \cite{Gubser}. In literature, various holographic superconductors
have been studied depending on the gravity theory in the bulk , from  Einstein theory \cite{Herzog},  Gauss-Bonnet \cite{GB}, Horava-Lifshitz theory \cite{HL} or by using non linear Born-Infeld electrodynamics \cite{born},lower dimensional models \cite{btz}, with conformal invariance  Weyl corrections \cite{weyl}, magnetic fields \cite{wen} and more \cite{Zhao:2013pk}-\cite{Pan:2012jf}. Also the Lifshitz black holes provide good background for holographic superconductors \cite{lifshitz}. Now it is interesting if one study the
holographic superconductors for a generic $\mathcal{F}(R)$ gravity. The higher order curvature corrections are very popular, because  some of these corrections  naturally appear in  string theory (corresponds to the finite 't Hooft coupling regime). For example the Gauss-Bonnet terms are the leading order terms in the effective action or the Weyl's tensor corrections are the conformal invariance part of the gravitational sector of a typical effective action. So, higher order curvature terms are important in extensions of effective models based on AdS/CFT. Briefly we review the motivation for $\mathcal{F}(R)$  modified gravity models.
 The first non trivial extension of the Einstein-Hilbert (EH) action is written as:
\begin{eqnarray}
S_{eff}=\int{d^4x \sqrt{-g}(\frac{R}{2\kappa^2})}-\int{d^4x \sqrt{-g}\Big[\gamma R^2+\gamma' R^{-1}+...\Big]}
\end{eqnarray}
It has many applications in cosmology and gravity. The first quadratic term used in inflationary scenarios \cite{starbonisky}. This term dominates in the regime  $R>>R_0$ where $R_0$ is a scale for the beginning of the decoupling of the gravity from the matter part of the action. The inverse term $R^{-1}$ can be estimated from the secular solar bounds \cite{1/R}. This term  dominates in small curvature regions. The idea of replacing the linear $R$ Einstein-Hilbert action action with a generic arbitrary function of $R$ is an old idea \cite{buchdahl}.  The first trivial attempt of modified
general relativity (GR) is to substitue Ricci scalar, $R\rightarrow\mathcal{F}(R)$ gravity \cite{Capozziello}. For this kind of geometrical modifications, we suppose
that the gravitational action  contains some higher order curvature terms
are growing with decreasing curvature. The final destination of these terms will be a late
time acceleration epoch. The resulting field equations are fourth order and has some meaningful cosmological implications. For example they predict the acceleration expansion of the inverse without any additional exotic dark component with extra degrees of the freedom. The $\mathcal{F}(R)$ action is locally Lorentz invariance. So this curvature correction is physically reasonable extension based on the geometrical objects. So such theory with Lorentz invariance can be used for a gauge/gravity duality via AdS/CFT conjecture. Thus from the CFT point of view there is no limitation to have a modified $\mathcal{F}(R)$ set up for holographic superconductors. The problem is how the non linear terms make the condensation harder or weaker. For example in the Gauss-Bonnet holographic superconductors if we investigate the system numerically or analytically we find that the coupling constant makes the condensation harder \cite{GB}. Also in the Weyl corrected holographic superconductors the same phenomena is observed \cite{weyl}. So, it seems that higher corrections in the bulk gravity sector affect the superconductivity on the quantum theory that is described on boundary as well by CFT. If the curvature correction provide for us some information about the physics of the superconductivity, it's natural that we include them in the action,specially with some non linear terms with a general form $\mathcal{F}(R)$. It's exactly which we did in this paper. Further for the scaling symmetry breaking part of the theory by the gauge field, we add a non linear term of the $F^{1/2}$ to the usual Maxwell Lagrangian. As we will show later, this term plays the role of the scaling invariance breaking of the system. The dual quantities on the boundary must be scale invariant, we want to specify the behavior of the dual quantities on the boundary by investigation of the effects of the  non linear Maxwell term $F^{1/2}$  in the probe limit.  The assumption about the probe limit is just as the first order approximation for treatment of the system by the perturbation theory and be taking the metric of the gravity part unperturbed. For full  description of the system it is needed to solve the highly non linear coupled system of the field equations of $\mathcal{F}(R)-\mathcal{L}(F)$ system. This system can be solved as the localized backreacted system. It can be the possible perspective project. But here we just focus on the probe limit in which we ignore from the backreaction of the matter fields on the background black hole solution in AdS metric. In brief,in this paper we would like to consider a $\mathcal{F}(R)$ gravity  coupled
with a scalar field $\psi$ and  a $U(1)$
 non linear electromagnetic field. We have discussed a type of solutions which also used recently in non linear effects of holographic superconductors \cite{Roychowdhury:2012vj}. We will take this exact black hole solution as the gravity part and we will study the scalar condensation on the dual theory using the gauge/gravity duality.\\
 %%%%%%%%%%%%%%%%%%%%%%%%%%%%%%%%%%%%%%%%%%%%%%%%%%%%%%%%%%%%%%%%%%%%%%%%%%%
The plan of this work is as following: In section 2 we have presented exact planar black hole solution in $\mathcal{F}(R)$ gravity . In section 3
we have formulated the basic set up for scalar condensation. In section 4 we will discuss the analytical matching solutions. In section 5, we discuss the holographic renormalization of the model. In sections 6,7,8 we have computed the critical temperature and  condensation parameter of our model and shown the
behavior of the it as a function of  temperature and the non linearity parameter. Further we will do a numerical estimation. We have concluded our results in the last section.

%%%%%%%%%%%%%%%%%%
\section{ $AdS_4$ solution in  $\mathcal{F}(R)$ gravity }
%%%%%%%%%%%%%%%%%%%%%
We take  the following action for a $\mathcal{F}(R)$ gravity plus a non linear Maxwell field $A_\mu$ and a massive scalar field $\psi$,
\begin{eqnarray}
 S=\int d^4x
\sqrt{-g}\Big(\frac{\mathcal{F}(R)}{2\kappa}+\mathcal{L}(F)-\Big[(D_{\mu}\psi)(D^{\mu}\psi)^{*}-m^2|\psi|^2\Big]\Big)\label{action} 
\end{eqnarray}
By definition $D_{\mu}=\partial_{\mu}-ieA_{\mu}$. Also electromagnetic part is
\begin{eqnarray}
\mathcal{L}(F)=-\frac{1}{4\pi}(F+2\beta\sqrt{-F})
\end{eqnarray}
Where $F=F^{\mu\nu}F_{\mu\nu}$ and $F_{\mu\nu}=\partial_{\mu}A_{\nu}-\partial_{\nu}A_{\mu}$. The full system of equation of motion reads
\begin{eqnarray}
\mathcal{F}_R R_{\mu}^{\nu}+\Big(\Box \mathcal{F}_R-\frac{1}{2}\mathcal{F}\Big)\delta_{\mu}^{\nu}-\nabla^{\nu}\nabla_{\mu}\mathcal{F}_R=\kappa T_{\mu}^{\nu}\label{geom}\\
\Box\psi=iA^{\mu}\partial_{\mu}\psi+i(-g)^{-1/2}\partial_{\mu}(\sqrt{-g}A^{\mu}\psi)+A^2\psi+m^2\psi\label{psieom}\\
-\frac{2}{\sqrt{-g}}\partial_{\mu}\Big[\frac{\delta \mathcal{L}(F)}{\delta F_{\mu\nu}}\Big]=i(\psi^{*}\partial^{\nu}\psi-\psi(\partial^{\nu}\psi)^*)+2A^\nu|\psi|^2\label{Feom}
\end{eqnarray}
Here the energy momentum tensor $T_{\mu\nu}$ is \cite{epjc},
\begin{eqnarray}
T_{\mu\nu}=\frac{F}{4\pi}diag\Big[1,1,\frac{2\beta}{\sqrt{-F}}-1,\frac{2\beta}{\sqrt{-F}}-1\Big]+(D_{\mu}\psi)(D_{\nu}\psi)^{*}+m^2|\psi|^2g_{\mu\nu}.
\end{eqnarray}
 The system composed of a generalized Einstein ,  Klein-Gordon and  Maxwell's equations.\\
Trace of (\ref{geom}) gives us
\begin{eqnarray}
R\mathcal{F}_R+3\Box \mathcal{F}_R-2\mathcal{F}=\kappa T \label{fT}
\end{eqnarray}
We need a black hole in $AdS_4$ background for bulk with constant curvature $R$, and in the probe limit, i.e. when $\kappa=0$. From (\ref{fT}) by assuming that $R=R_0$ we obtain
\begin{eqnarray}
R_0 \mathcal{F}'(R_0)=2\mathcal{F}(R_0)
\end{eqnarray}
Now we take\cite{epjc},
\begin{eqnarray}
\mathcal{F}(R)=R+\frac{6}{l^2}+2\alpha\sqrt{R+R_0}
\end{eqnarray}
So, we obtain
\begin{eqnarray}
\alpha=-\frac{\sqrt{2}}{7}\Big[\sqrt{R_0}+\frac{12}{l^2\sqrt{R_0}}\Big]
\end{eqnarray}
So, (\ref{geom}) in absence of electromagnetic and scalar fields, and by set $\kappa=0$ has the following exact planar solution
\begin{eqnarray}
ds^2=-f(r)dt^2+\frac{dr^2}{f(r)}+r^2(dx^2+dy^2),\ \ f(r)=\frac{r^2}{l^2}(1-\frac{r_{+}^3}{r^3})\label{g}
\end{eqnarray}
Here horizon size $r_+$ is related to the mass and thought it,$r_+$ is a function of $\alpha,R_0$.
 Direct substitution of  (\ref{g}) in vacuum case of (\ref{geom}) for $(r,r)$ component, when $T_{\mu\nu}=0$ we obtain
\begin{eqnarray}
r_{+}^3=\frac{3M\mathcal{F}'(R_0)}{\mathcal{F}(R_0)}=\,{\frac {24M{l}^{2}}{7\,\sqrt {2}\alpha\, \left( -\frac{7}{4}\,\alpha\,
\sqrt {2}l+\frac{1}{4}\,\sqrt {98\,{\alpha}^{2}{l}^{2}-192} \right) l+24}}
\end{eqnarray}
If we study Einstein case $\alpha=0$, the above expression recovers $r_{+}^3=Ml^2$. The Hawking-Bekenstein
temperature of the horizon calculated by the killing vectors on horizon , $r=r_{+}$ reads as
$$
T_{BH}=\frac{f'(r_{+})}{4\pi}=\frac{3r_{+}}{4\pi l^2}
$$
This temperature coincides on the temperature of the  dual gauge field theory on the boundary via CFT. In the following section we set up the model for holographic superconductor based on the bulk metric given by (\ref{g}).
%%%%%%%%%%%%%%%%%%%%%%%%%%%%%%%%%%%%%%%%%%%%%%%%%%%%%%%%%%%%
\section{Field equations for scalar condensate}
%%%%%%%%%%%%%%%%%%%%%%%%%%%%%%%%%%%%%%%%%%%%%%%%%%%%%%%%%%%%
Back to the model of $\mathcal{F}(R)$ gravity with non linear Maxwell field and scalar, presented in (\ref{action}). Note that when $\beta\rightarrow 0$, $\alpha=0$ it's
just the Einstein-Maxwell model with $\Lambda<0$ and $l$ denotes the AdS radius. If we put $\alpha=0,\ \ \beta=0$, the model describes usual holographic superconductors with Einstein-Hilbert gravity in the bulk\cite{Hartnoll}. We are interesting to investigate the phase transitions in the hairy background with $\alpha,\beta\neq0$.
 Let us  firstly to consider the case of the usual holographic super conductors (HSC), i.e. in the probe limit of a Einstein-Hilbert action with Maxwell field and coupled minimally to a complex  charged scalar field with the following action
\be
 S=\int d^4x
\sqrt{-g}\Big[\frac{R}{2\kappa}+\Lambda_{eff}-\frac{1}{4}F_{\mu\nu}F^{\mu\nu}-|\partial
\psi-ieA\psi|^2-m^2|\psi|^2\Big]\label{s}
\ee
Here $\Lambda_{eff}$ denotes the effective negative cosmological constant in the AdS space time  and $e$ is the electric charge, $A_{\mu}dx^{\mu}=\phi(r)dt$. We assume that the asymptotically $AdS_4$ background metric is
\be
g_{\mu\nu}=diag\Big(-g(r)e^{-\chi(r)},\frac{1}{g(r)},r^2,r^2\Big)
\ee
From boundary conditions on the horizon $r_{+}$, since $g(r_{+})=0$, to avoid from the divergence in the norm of the $U(1)$ gauge field $A_{\mu}$, we impose $\phi(r_{+})=0$. Now the independent parameters evaluated on horizon $r=r_{+}$ given by  $\{r_{+},\psi(r_{+}),\phi'(r_{+}),\chi(r_{+})\}$. The system of the equations are invariance under scaling symmetries
\be
r\rightarrow ar,\ \ (x,y,t)\rightarrow (x,y,t)/a,\ \ g\rightarrow a^2 g,\ \ \phi \rightarrow a\phi
\ee
Further the system is scale invariance under an additional scaling symmetry
$$
e^{\chi(r)}\rightarrow a^2 e^{\chi(r)},\ \ t\rightarrow at,\ \ \phi\rightarrow\phi/a
$$
Now if we write the field equations of the modified non linear model (\ref{s}), it is easy to show that the new parameter $\beta$ breaks the scaling invariance of the model.\\
Why we need to breaking the scale invariance?. It's a basic question which it has been answered previously \cite{breaking}. Indeed in a gauge theory when the spontaneous symmetry breaking of scale invariance happens, it naturally leads to \emph{the confinement of static electric charges in Coulomb interaction}. Such term creates a new linear modification to the Coulomb potential, and as 't Hooft has showed that \emph{such confinement can be addressed to  a linear first order term in the dielectric field $D$ as the electrical response function of the system} \cite{thooft}. This linear modification of the Coulomb  potential needs a local counter term
in the Lagrangian that renormalize the infrared divergence in the Coulomb
potential. The detail of this holographic renormalization is beyond the scope of our work(See for example \cite{holo-ren}). We are not able  to explain the method of such renormalization to cancel this boundary divergence term in this paper \footnote{This part is in progress.}. A brief explanation of the problem will be presented in section V. Also in the next sections we will show that, when we are working on the AdS boundary $r\rightarrow\infty$, this linear term $\phi\propto r$ introduces a divergence term appear in the form of  $\beta r$, which is a natural reason for that we label $\beta$ as a scale invariance breaking parameter.\\
Following \cite{breaking}, now we want to expose the role of the non linearity of the Maxwell strength field\footnote{Such simple model ends finally to confinement,and to string solutions. $\beta\sqrt{-F}$ in the term due to the spontaneous symmetry breaking of scale invariance of our proposed model. But now we working with the $U(1)$ gauge fields not $SU(2)$ which has been discussed previously in \cite{breaking}. }.\\
 Under a scale transformation $x\rightarrow \lambda x$, the quantity  $\sqrt{-F(x)}$  transforms as
\be
\sqrt{-F(x)}\rightarrow \lambda^2 \sqrt{-F(\lambda x)}\label{scale}
\ee
It is adequate to demonstrate a new auxiliary field as a one-form $\omega=\sqrt{-F}$, here $\omega$ is an elementary field:
\be
\omega=\epsilon^{\mu\nu}\partial_{[\mu}A_{\nu]}
\ee
From the field equation for gauge field $A_{\mu}$ we have
\be
\omega=\sqrt{-F(x)}+\Theta
\ee
Here $\Theta$ is the integration constant which it  spontaneously breaks the scale invariance. The reason is that   both $  \omega,\ \ \sqrt{-F(x)}$ transform in a similar form according to the transformation law Eq.(\ref{scale}), but $\Theta$ does not transform. It's obvious to show  that  dimension of $\Theta\cong F\cong \text{length}^{-2}\sim\beta^2$. So this is another evidence to taking $\beta$ as a scale invariance breaking coupling parameter in our model.\\
Back to our model with the previous physical discussions,we want to discuss the black hole solutions of the (\ref{s}) in the probe limit.
The parameter $\beta$ as a "` \emph{spontaneous symmetry breaking for the the scale invariance}"'   plays the role of the effective string constant in the linear modification of the Coulomb potential \cite{thooft} but for our holographic picture and using field theory, it defines the  dielectric field $D$ and it dominates in regime
$|D|<<|D_0|$ and the asymptotic limit of the scalar potential $\phi$ on the AdS boundary $r\rightarrow\infty$. 
 In this paper we are interesting to investigate  the effects of the $\beta$ on the critical
  temperature $T_c$ and the condensation parameter $<O_{\Delta_{\pm}}>$. At least we want to describe the condensation phase for small values of the $\beta$ of order $O(\beta)$.
In the probe limit by fix the metric as given by (\ref{g}), we ignore from the back reaction of all fields. By adopting a static gauge we have
\be
 A_{\mu}=\phi(r)\delta_{\mu t},\ \
\psi=\psi(r) \ee
 Explicitly
  \be
   F=-\frac{1}{2}\phi'^2,\ \ |\partial
\psi-iA\psi|^2=f \psi'^2-\frac{(\psi\phi)^2}{f}
 \ee
 The equations of motion for $e=1$ read
\begin{eqnarray}
\phi''+\frac{2}{r}\phi' =\frac{2\beta\sqrt{2}}{r}+\frac{8\pi \psi^2}{f}\phi\label{phir}\\
\psi''+\Big(\frac{f'}{f}+\frac{2}{r}\Big)\psi'+\frac{\phi^2}{f^2}\psi=\frac{m^2}{f}\psi\label{psir}
\end{eqnarray}
When $\beta=0$ the equations reduce to the \cite{Hartnoll}. In next section , following recently analytical  solving method for the equations of motion near the critical point, we will describe the full phase of system.
%%%%%%%%%%%%%%%%%%%%%%%%%%%%%%
\section{Solution of the field equations }
%%%%%%%%%%%%%%%%%%%%%%%%%%%%%%%%%%%%%%%%%%%%%
The equations of the motion which have been obtained in the previous section are highly coupled non linear equations which can not be solved analytically. So, for convince the numerical algorithms are preferred. The common numerical scheme is the shooting method. So, it is suitable if we can solve the Eqs. by applying another semi analytical method. One of the best tools is the matching method proposed in \cite{Gregory:2009fj} and recently motivated by several authors\cite{Zhao:2012kp}-\cite{Roychowdhury:2012hp}. In matching method we connect smothly the solutions nearly the AdS boundary
$z=0$ to the neighbor of the horizon solutions $z=1$ in a mid point $0<z_m<1$.
Usually it takes $z_m=\frac{1}{2}$. The matching method can not predict the correct behavior of the fields. As the authors showed before \cite{Zhao:2012kp}-\cite{Roychowdhury:2012hp}, the matching method is  a good approximation in the comparison to the numerical solutions. For example the matching method gives us a reasonable behavior near the AdS horizon . So, this semi analytical method is useful because it's simplicity in application. So we will follow it.
 First we must write the
solutions for these two different regions.

\subsection{Solutions near the horizon $z=1$}
We expand field functions $\psi(z),\phi(z)$ in a Taylor series nearby the point $z=1$:
\begin{eqnarray}
\phi(z)=\phi(1)-\phi'(1)(1-z)+\frac{1}{2}\phi''(1)(1-z)^2+...\\
\psi(z)=\psi(1)-\psi'(1)(1-z)+\frac{1}{2}\psi''(1)(1-z)^2+...
\end{eqnarray}
Since  that on horizon the gauge field $A_{\mu}$ must be finite, we
impose that $\phi(1)=0$,$\phi'(1)<0,\psi(1)>0$  to preserve $\{\phi(1),\psi(1)\}>0$. Now we
rewrite the  equations (\ref{phir}), (\ref{psir}) in the following
equivalent form of the coordinate $z$
\begin{eqnarray}
z^4 \phi''-(8\pi r_{+}^2)\frac{\psi^2}{f}\phi=2\beta\sqrt{2}r_{+} z\label{phiz}\\
\psi''+\frac{f'}{f}\psi'-(\frac{r_{+}}{z^2})^2(-\frac{m^2}{f}+\frac{\phi^2}{f^2})\psi=0\label{psiz}
\end{eqnarray}
In limit $\beta=0$, the Eqs.
(\ref{phiz}), (\ref{psiz}) reduce to the usual field equations in linear Maxwell theory in unit $8\pi G\equiv 2,\ \ G\equiv1$ \cite{Hartnoll}.
Our goal here is to study  effect of the non linearity $\beta$
for superconductivity. Further we want to study the effect
of  $\mathcal{F}(R)$ parameter $\alpha$ in the critical temperature $T_c$.
Expanding (\ref{phiz}) near $z=1$ we obtain
\begin{eqnarray}
\phi''(1)=2\beta\sqrt{2}r_{+} +(8\pi
r_{+}^2)\frac{\psi^2(1)}{f'(1)}\phi'(1),
\end{eqnarray}
here we taken limit from the term $\frac{\phi(z)}{f(z)}$ in $z=1$.
So we have the following expression , which is valid only near $z=1$,
\begin{eqnarray}
\phi(z)=b \left( 1-z \right) + \left( \beta\,\sqrt {2}r+{\frac {r_{+}{a}^{2}b}{T}}
 \right)  \left( 1-z \right) ^{2}
\label{phiaprox}
\end{eqnarray}
Similarly, using the (\ref{psiz}) and by expansion near $z=1$ we
have
 the following expression is valid as an approximate solution for $\psi(z)$ only near $z=1$
\begin{eqnarray}
\psi(z)=a \left( 1-\frac{1}{4}\,{\frac {{m}^{2}r \left( 1-z \right) }{\pi \,T}}+\frac{r_{+}^{2}}{4}
 \left( -\frac{1}{16}\,{\frac {{b}^{2}}{{\pi }^{2}{r_{+}}^{2}{T}^{2}}}-\frac{1}{16}
\,{\frac {{m}^{2}}{{\pi }^{2}{T}^{2}}}+{\frac {64}{9}}\,{\frac {{m}^{2
}{\pi }^{2}{l}^{4}{T}^{2}}{{r_{+}}^{2}}} \right)  \left( 1-z \right) ^{2}
 \right) 
\label{psiaprox}
\end{eqnarray}

%%%%%%%%%%%%%%%%%%%%%%%%%%%%%%%%%%%%%%%%%%%%%%%%%%%%
\subsection{Solutions near AdS boundary $z=0$}
%%%%%%%%%%%%%%%%%%%%%%%%%%%%%%%%%%%%%%%%%%%%%%%%%%%%%
In the asymptotic AdS boundary, as we know the following solutions are valid
\begin{eqnarray}
\psi=D_{+}z^{\Delta_{+}}+D_{-}z^{\Delta_{-}}\label{psiasympt}\\
\phi=\frac{\beta\sqrt{2}}{z}r_{+}^3+\mu-\rho z\label{phiasympt}
\end{eqnarray}
here $\mu,\rho$ correspond to the
chemical potential and charge density in the dual theory
respectively. The Eq. (\ref{phiasympt}) gives the behavior of the scalar field at the vicinity of the AdS  boundary $(z=0)$ . However, the first term that is proportional to $\frac{\beta}{z}$  seems quite strange for us. This term is divergent at $z=0$. This suggests that we need to renormalize the divergence by introducing an appropriate holographic counter term(s). As we mentioned before the renormalization via holographic tools is needed for such counter terms, but we are not able here to discuss and clarify it . In fact, the details are so far from our paper. We return  to our model to show that why in our model for holographic superconductors the non-optimal term $\frac{1}{z}$ appears. The problem here is similar to finding   the static corected Columb potential of a pair of  heavy quark-antiquark pair  when it is nedded to modify Columb interaction  in the frame work of a non-Abelian gauge theory.  However, the inter-quark potential is governed by the geometry. In other words, the inter-quark potential within the probe approximation is governed by the physics of gluons. Then, it seems that,  it is difficult ,   within the probe approximation, to induce the linear potential by switching on $\beta$. But, we will show that such term, appears in the $\mathcal{F}(R)$ set up for holographic superconductors , even in the probe limit. Actually, it seems that the role of the geometry in the the inter-quark potential plays by the infinite numbers of the power curvature terms like $R^m,\ \ m\in Z$, which are obtained from the $\sqrt{R+R_0}$.\footnote{ Such calculation
have been done using the \emph{gauge-invariant},\emph{ path-dependent},
\emph{variables method} \cite{quark} which is in agreement with the \emph{'t Hooft perturbative
treatment} for achieving confinement}.
 At the boundary $\mu$ has mass dimension one and $\rho=\frac{\mu}{r_{+}}$ is of mass dimension two and $<O_{\pm}>$ denote expectation values of dual fields. Near the AdS boundary, $z=0$, with conformal dimension
 \be
\Delta_{\pm}=\frac{3}{2}\pm\sqrt{m^2-m_{BF}^2},\ \ m_{BF}^2=-\frac{9}{4},
\ee
 Since we are interesting in both of these falloffs be renormalizable, and further for stability reasons we take:
\begin{eqnarray}
D_{+}=0,\ \  <O_{-}>=\sqrt{2}D_{-}\label{q1}
\\
D_{-}=0,\ \  <O_{+}>=\sqrt{2}D_{+}\label{q2}
\end{eqnarray}
With the normalization factor $\sqrt{2}$ . The two Eqs.(\ref{q1},\ref{q2}) correspond to two alternate choices of quantization. Both of the terms with conformal dimensions $\Delta_{\pm}$ fall off and we can keep they.  In conclusion, the quantization scheme is a valid procedure. The scalar field is asymptotic
to  $<\cal{O_{\pm}}>$ and these are dual to  operators with dimension $\Delta_{\pm}$. In fact,it is possible to write this quantization scheme in terms of other parameters as it has been proposed Ref.~\cite{jpa} for case of Lifshitz black holes. However, in this work  we restrict ourselves to the fall off with $\Delta_{+}$. It means we choice the conformal dimension and consequently we fix quantum operator on the boundary. Even if we assume that there exists a specific combination of the these two operators, then the scalar field with both of them is not normalizable. So, either (\ref{q1}) or (\ref{q2}) holds, but not both. We will set $D_{-}=0$.
 In next section by matching (\ref{phiaprox},\ref{psiaprox}) with the
(\ref{psiasympt},\ref{phiasympt}) with the matching point $z_m=\frac{1}{2}$, (this is independence from the choice of the $z_m$) we will study the $T_c$.

%%%%%%%%%%%%%%%%%%%%%%%%%%%%%%%%%%%%5
\section{On holographic renormalization of  $\mathcal{F}(R)$}
%%%%%%%%%%%%%%%%%%%%%%%%%%%%%%%%%%%%%%%%%%%%
Holographic renormalization in Einstein gravity is a well studied topic (see for example \cite{holo-ren}). But in extended models of gravity, like $\mathcal{F}(R_{\mu}^{\nu})$ it is a new problem in progress. We know that for a model $\mathcal{F}(R_{\mu}^{\nu})$ such possibility exists at least in three dimensions \cite{Loran:2013fca}. The technique is how we cancel the divergence terms on the on-shell action using a "`\emph{non-covariant cut off independent term}"'. We mention here some published results on holographic renormalization in $\mathcal{F}(R)$. One of the main problems deal with in CFT is how we identify the expectation value of the traceless tensor of CFT to a correspondence quantity (of course we mean another energy momentum tensor quantity) in weak gravity bulk action. The quick answer is the correspondence exists between Brown-York tensor and CFT one \cite{holo-ren}. It needs to identify the surface term of action by appropriate boundary condition. This surface term as we know for Einstein gravity is Gibbons-Hawking counter term. But in $\mathcal{F}(R)$ gravity the situation is completely different. As we know $\mathcal{F}(R)$ in Jordan frame reduces to a sub class of Brans-Dicke models \cite{Nojiri:2006gh}. It is very interesting that there is not exist any physically acceptable counter term for $\mathcal{F}(R)$ gravity to identify to CFT dual quantity \cite{madsen}. It is a remarkable result that the appropriate boundary condition for our $\mathcal{F}(R)$ model is to set the variation of the curvature $\delta R=0$ on boundary.  . So, as we mentioned before we are not able to perform a holographic renormalization on our model to cancel the divergence term of linear $r$ or $z^{-1}$ on the conformal boundary. It remains as an open problem for any holographic study of $\mathcal{F}(R)$ models.

%%%%%%%%%%%%%%%%%%%%%%%%%%%%%%%%%%%%%%%%%%%%%%%%%%%%
\section{Calculating $D_{+}=\sqrt{2}<O_{+}>,T_c$}
%%%%%%%%%%%%%%%%%%%%%%%%%%%%%%%%%%%%%%%%%%%%%%%%%%%%%%%%
By logarithmic continuity $\frac{\zeta'}{\zeta}|{z_m= \frac{1}{2}},\ \ \zeta=\{\phi,\psi\}$, we obtain the following algebraic equations (taking $\psi(1)=a,-\phi'(1)=b,\ \ (a,b>0),\ \  f'(1)=-4\pi r_{+} T<0,\ \ f''(1)={\frac {32}{3}}\,{\pi }^{2}{l}^{2}{T}^{2}
$)
\begin{eqnarray}
2\,\beta\,\sqrt {2}r+\mu-\frac{q}{2}=\frac{b}{2}+\frac{\beta\sqrt {2}r_{+}}{4}+\frac{1}{4}\,{
\frac {r_{+}{a}^{2}b}{T}}
\label{eq1}\\
-4\,\beta\,\sqrt {2}r_{+}-q=-b-\beta\,\sqrt {2}r_{+}-{\frac {r_{+}{a}^{2}b}{T}}
\label{eq2}\\
 D_{+} \left( \frac{1}{2}\right) ^{\Delta_{+}}=a \left( 1-\frac{1}{8}\,{
\frac {{m}^{2}r}{\pi \,T}}+\frac{1}{16}\,{r}^{2} \left( -\frac{1}{16}\,{\frac {{b}^{2}
}{{\pi }^{2}{r}^{2}{T}^{2}}}-\frac{1}{16}\,{\frac {{m}^{2}}{{\pi }^{2}{T}^{2}}
}+{\frac {64}{9}}\,{\frac {{m}^{2}{\pi }^{2}{l}^{4}{T}^{2}}{{r}^{2}}}
 \right)  \right) 
\label{eq3}\\
2\, D_{+} \left(\frac{1}{2}\right) ^{\Delta_{+}}\Delta_{+}=a \left( \frac{1}{4}
\,{\frac {{m}^{2}r}{\pi \,T}}-\frac{1}{4}\,{r}^{2} \left( -\frac{1}{16}\,{\frac {{b}^{
2}}{{\pi }^{2}{r}^{2}{T}^{2}}}-\frac{1}{16}\,{\frac {{m}^{2}}{{\pi }^{2}{T}^{2
}}}+{\frac {64}{9}}\,{\frac {{m}^{2}{\pi }^{2}{l}^{4}{T}^{2}}{{r}^{2}}
} \right)  \right) 
\label{eq4}
\end{eqnarray}
After a simple calculation we obtain
\begin{eqnarray}
D_{+}=\mathcal{C}\sqrt{1-\frac{T}{T_c}}\label{D}
\end{eqnarray}
Here, the critical temperature,  $T_c=T_c(\beta)$. By more carefully study,
we detect  a linear relation between $\rho$ and $T_c$. For some positive constants $a$ and $b$  it is not possible for $T_c$ to go to zero for negative enough $\beta$, because always $\beta\geq 0$. So the case of the negative $T_c$ is absent here. The linear relation between the critical temperature and the charge density has been reported in the non relativistic model of gravity , for example in Horava-Lifshitz set up for holographic superconductors \cite{HL}. In non relativistic case, it arises from non relativistic nature of the Horava modification of the Einstein-Hilbert action as a critical anisotropic scaling of the space $x$, time $t$ coordinates. In Horava-Lifshitz gravity the total action  is power counting renormalizable (in the more precisely form "`Stochastic" renormalizable) at the critical exponent $z=3$. This feature is a common feature of the Horava-Lifshitz type of the holographic superconductors and the present model of $\mathcal{F}(R)$ set up for scalar condensation.  Further, we find the
value of $T_c$  grows up when $\beta$ increases because always $\rho>0$ and also $r_+>0$. Consqequently, a system with a positive non linear term has a more harder condensation
scheme. The phenomena here is very similar to the case, which it has been happen on holographic superconductors with  quasi-topological curvature corrections \cite{quasi}.

%%%%%%%%%%%%%%%%%%%%%%%%%%%%%%%%%%%%%%%%%%%%%%%%%%%%%%%%%%
\section{Properties of the critical temperature $T_c$}
%%%%%%%%%%%%%%%%%%%%%%%%%%%%%%%%%%%%%%%%%%%%%%%%%%%%%%%%
We denote by
\begin{equation}
T_c^0\propto\sqrt{\rho},\ \ \beta=0.
\end{equation}
 Note that in general case,
\be
T_c(\beta)=T_c^0+\beta \delta T_c
\ee
Since $\delta T_c\geq0$, thus as a first result, we observe that
\be
T_c(\beta)\geq T_c^0,\ \ \beta>0
\ee
It shows that when the non linearity increases, the critical temperature increases, and the condensation becomes harder.
Also,we observe that when the non linearity $\beta$ increases, the critical temperature has a higher peak and it causes the condensation harder.  Noting that  the temperature $T_c$ always remains  positive.

%%%%%%%%%%%%%%%%%%%%%%%%%%%%%%%%%%%%%%%%
\section{Properties of the $<O_{\Delta_{+}}>$}
%%%%%%%%%%%%%%%%%%%%%%%%%%%%%%%%%%%%%%%%
From equation (\ref{D}) we can calculate the $<O_{+}>=\frac{D_{+}}{\sqrt{2}}$. Explicitly we obtain
\begin{eqnarray}
<O_{+}>=\mathcal{C}\sqrt{1-\frac{T}{T_c}},\\
\ \  \mathcal{C}= \frac{2^{\Delta_{+}-1/2}}{\sqrt{\pi}}\Big(\frac{\Delta_{+}}{(\Delta_{+}+2)^3}\Big)^{1/4}\sqrt{\rho+3\sqrt{2}\beta r_{+}^3}.
\end{eqnarray}
We observe that $<O_{\Delta_{+}}>|_{T = T_c}=0$,  condensation
occurs for $T < T_c$. This phase transition is happen continously \cite{Hartnoll}, $<O_{\Delta_{+}}>\propto(T_c - T)^{1/2}$  coincides withthe result of mean field theory. This equation is valid only near the critical point $  T=T_c$. Specially, for small values of the non linearity parameter $\beta$ we have
\begin{eqnarray}
<O_{+}>^{\beta}\approx\Big(\mathcal{C}|_{\beta=0}+\beta \Big[\frac{\partial \mathcal{C}}{\partial \beta}\Big]_{\beta=0}\Big)\sqrt{1-\frac{T}{T_c}}
\end{eqnarray}
Again, we observe that in the regime of the small deviation from the linear electrodynamics,  $<O_{\Delta_{+}}>^{\beta}\propto<O_{\Delta_{+}}>^{0}$which is obtained by Maxwell field .

Additional comment on the role of $\alpha$ in condensation phase can be addressed , in which we studied the effect of $\alpha>0$ term in value of $T_c$. It shows that when $\mathcal{F}(R)$ coupling parameter $\alpha$ increses, also $T_c$ increases and it shows that condensation becomes more than before (when Einstein action used with $\alpha=0$) harder.

%%%%%%%%%%%%%%%%%%%%%
\section{conclusions}
%%%%%%%%%%%%%%%%%%%%%%%%
In summary, for the first time in literature \cite{pity},  we constructed  a holographic toy model for $\mathcal{F}(R)$ gravity as the gravity dual of a high temperature 
superconductor. As the gravity model, we probe a $AdS_4$ black hole as an exact solution for a generic type of $\mathcal{F}(R)$
 gravity with non linear Maxwell fields, and we have studied numerically and semi analytically, holographic superconductors properties like critical temperature and critical exponent.
We introduced the non linearity parameter $\beta$ as a parameter for spontaneous symmetry breaking on the scale invariance of the system in the presence of the non linear higher orders curvature terms, come from the $\sqrt{R+R_0}$.
 We compute analytical aproximated solution and we have found that there is also
a critical temperature like the relativistic (non-relativistic Horava-Lifshitz model) case,when $T<T_c$ the system undergoes a phase transition due to charged  condensation field.
We show that the scalar condensation in $f(R )$ theory with nonlinear $U(1)$ gauge field in four dimension and Gauss-Bonnet theory  have similar features.  The superconductor model has a similar pattern as quasi-topological gravity.

\newpage

\begin{figure*}[thbp]
\begin{tabular}{rl}
\includegraphics[width=7.5cm]{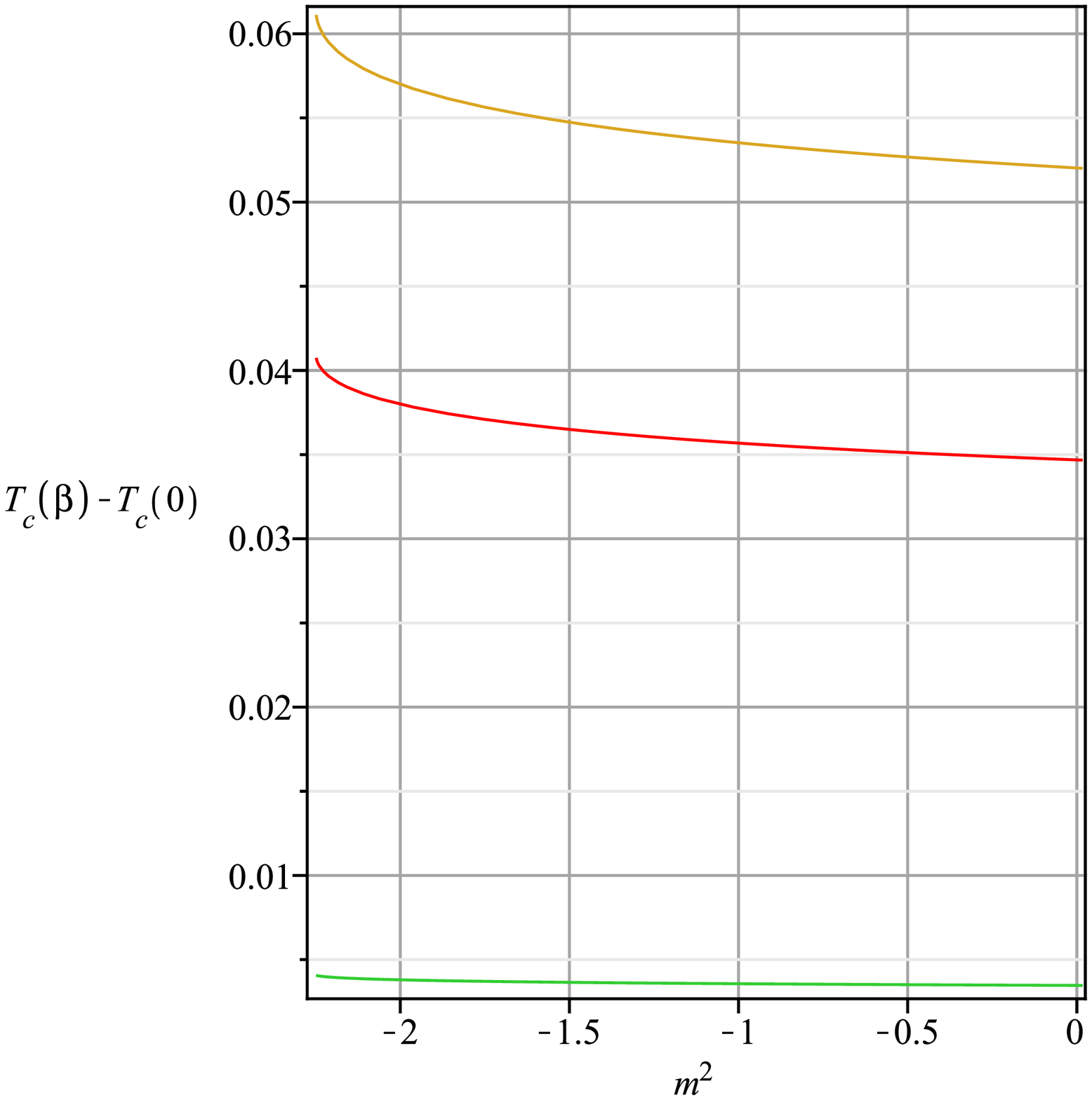}\\
\includegraphics[width=7.5cm]{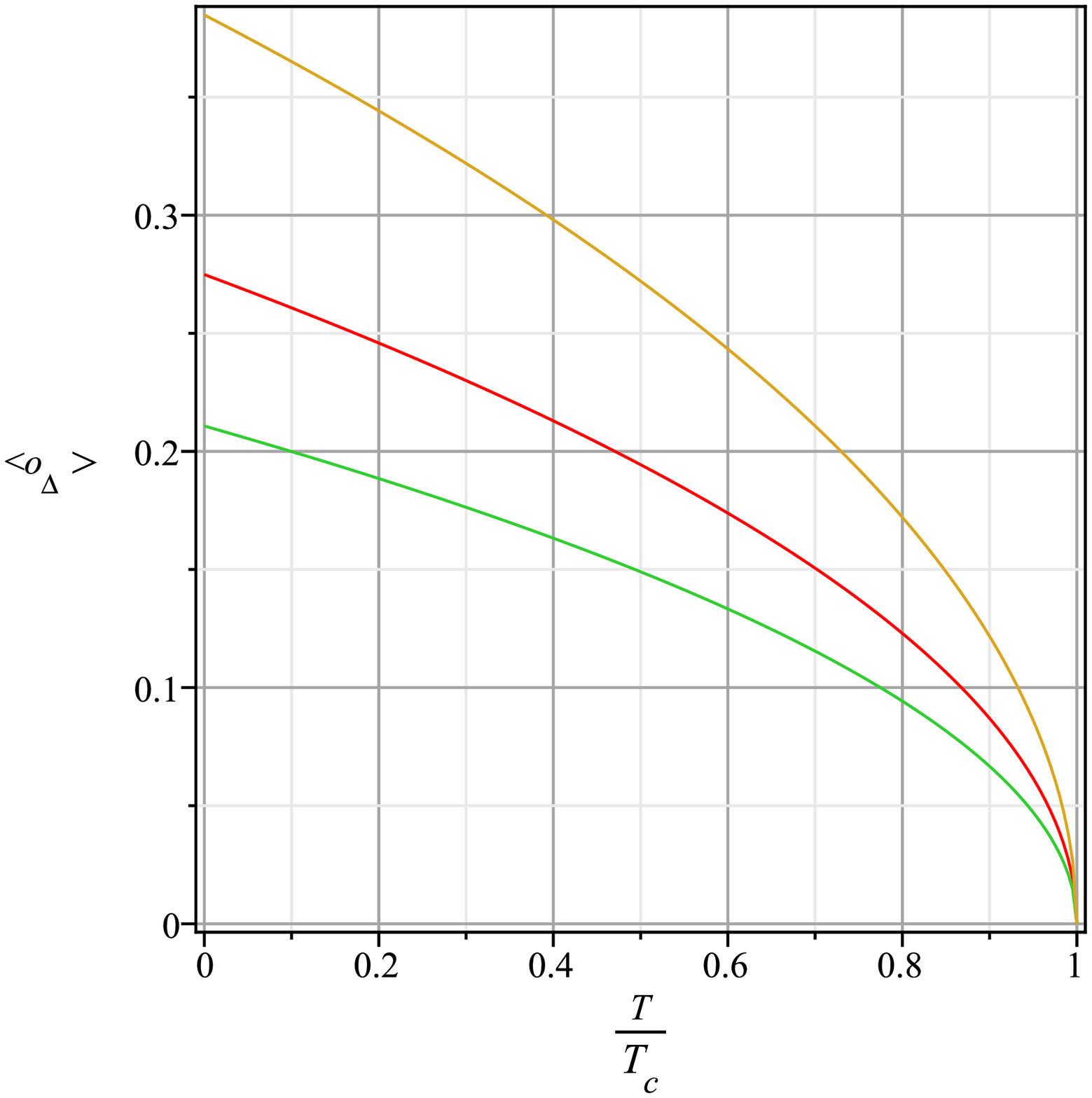} \\
\includegraphics[width=7cm]{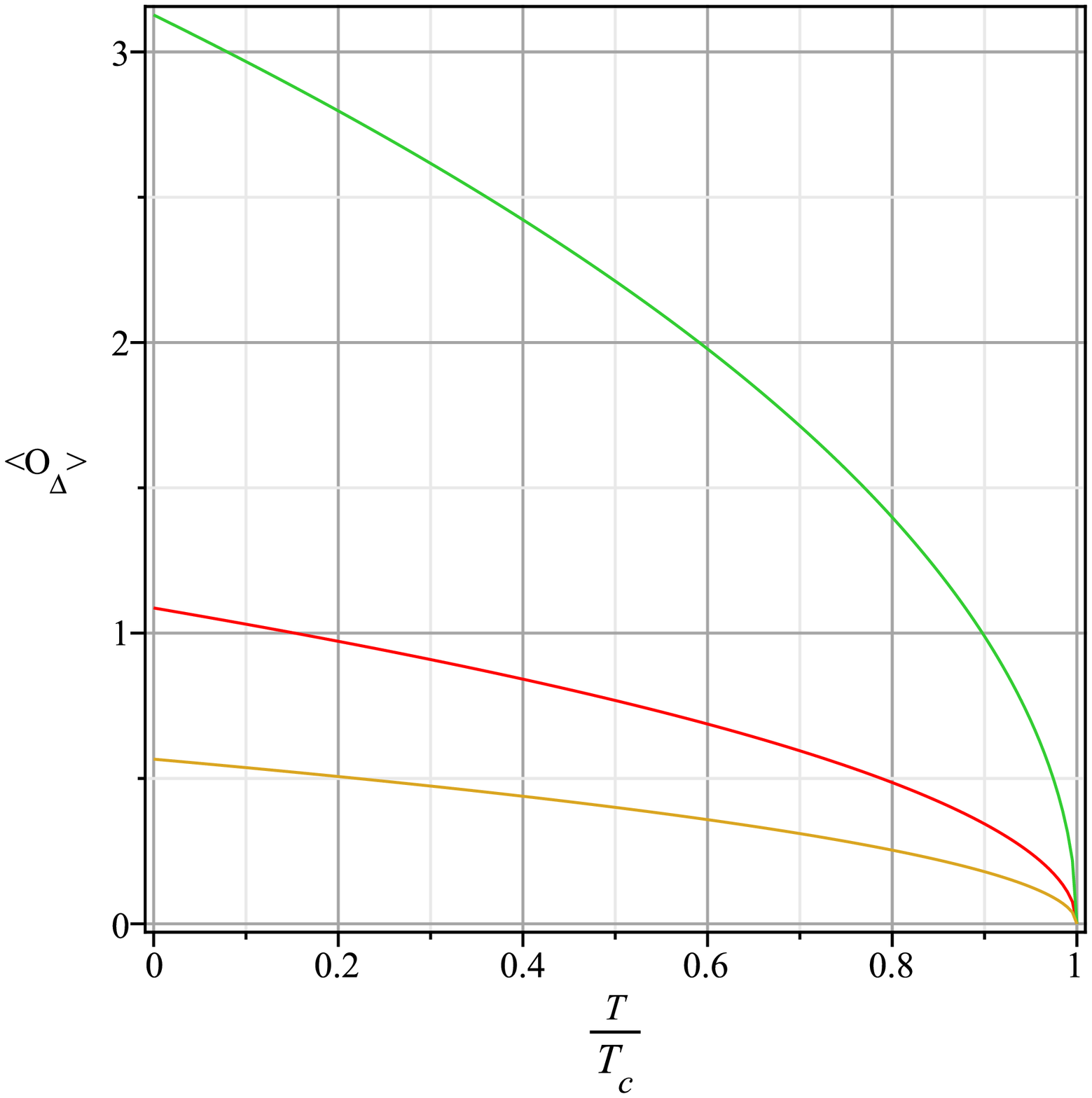}\\
\end{tabular}
\caption{ (\textit{Top })  Difference of $T_c(\beta)-T_c(0)$ vs. $m^2$ for different values of $\beta$ from top to bottom, by increasing $\beta$ it increses.   (\textit{Middle}) Condensation $<\mathcal{O}>$ vs. temperature for different $\beta$.  (\textit{Bottom }) Condensation $<\mathcal{O}>$ vs. temperature for values of $\alpha$. }
\end{figure*}

\begin{figure*}[thbp]
\begin{tabular}{rl}
\includegraphics[width=7cm]{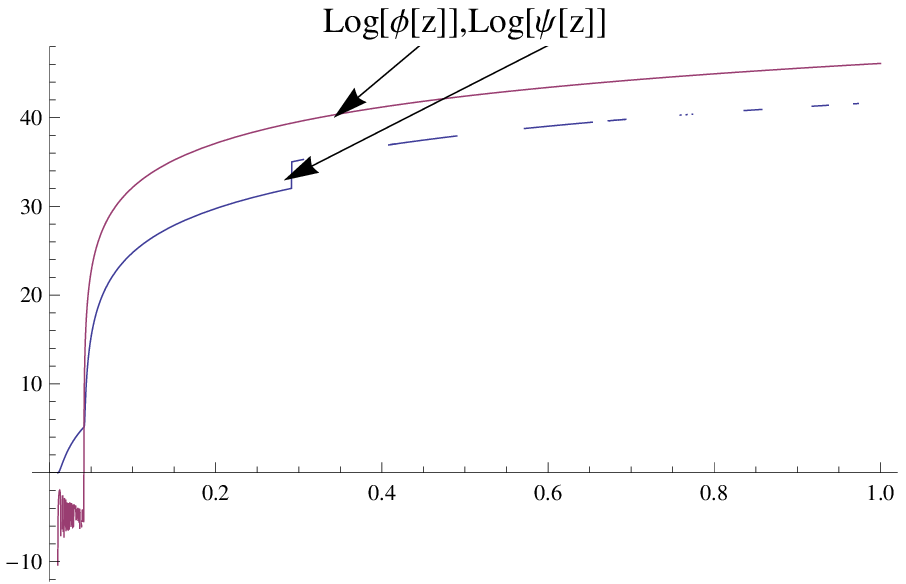} \\
\includegraphics[width=7cm]{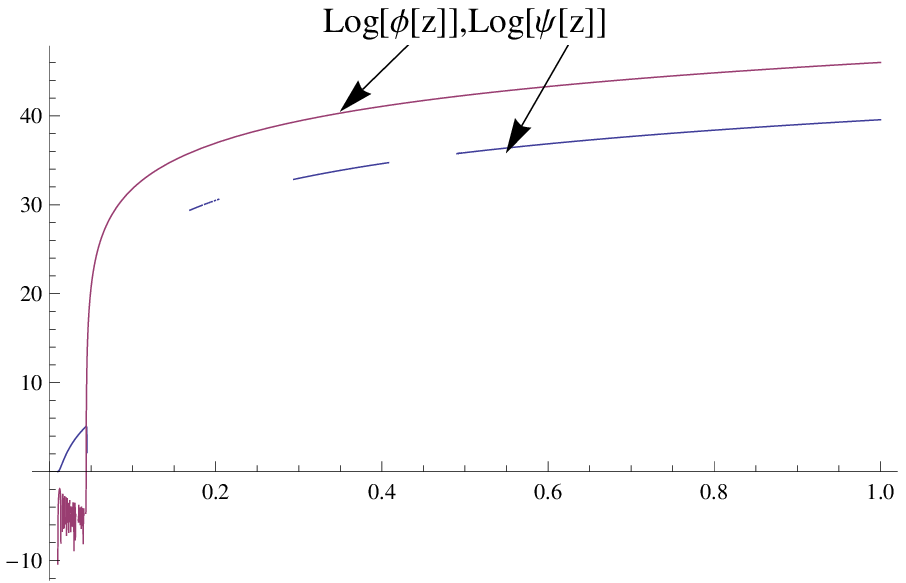}\\
\includegraphics[width=7cm]{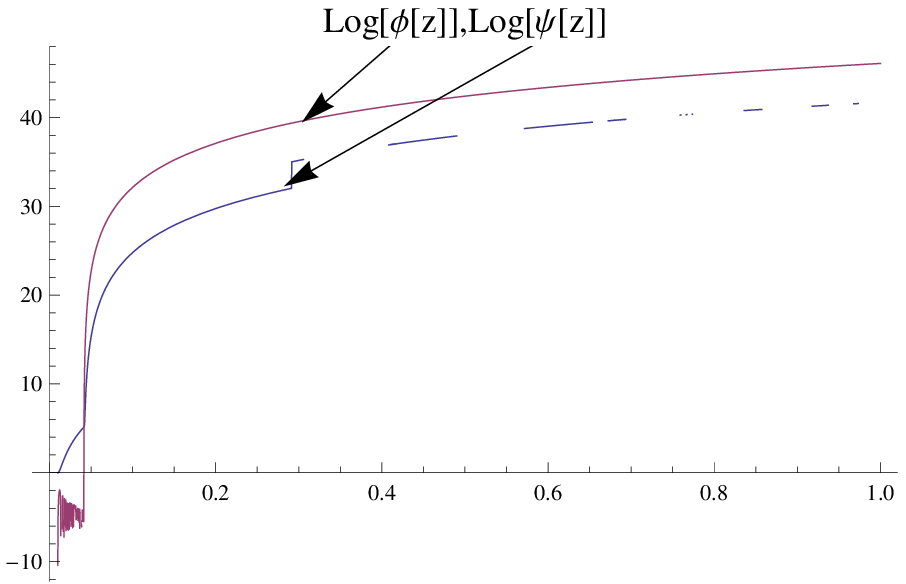} \\
\end{tabular}
\caption{ (\textit{ Left})  Field solutions  numerically for $m^2=-1,-2,-9/4$ from top to bottom. }
\end{figure*}

\begin{figure*}[thbp]
\begin{tabular}{rl}
\includegraphics[width=7.5cm]{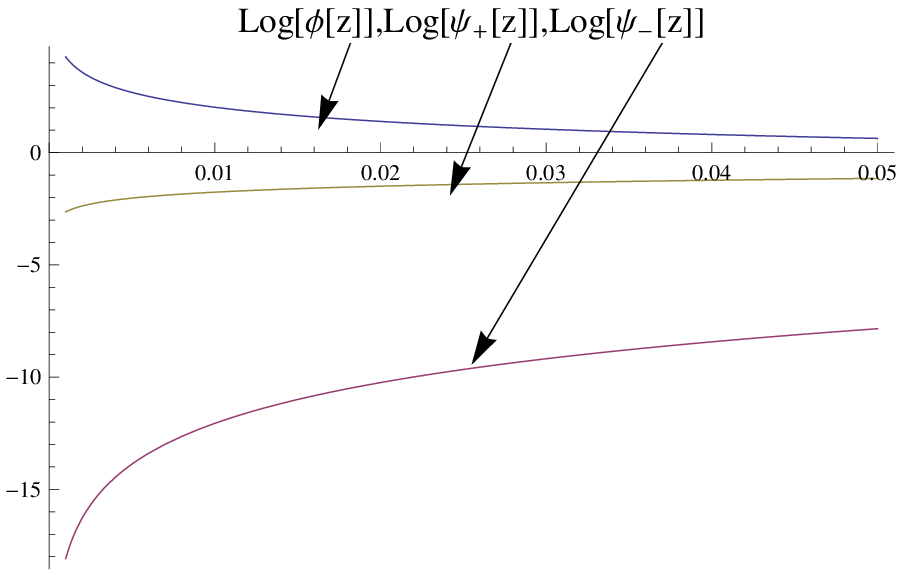}&
\includegraphics[width=7.5cm]{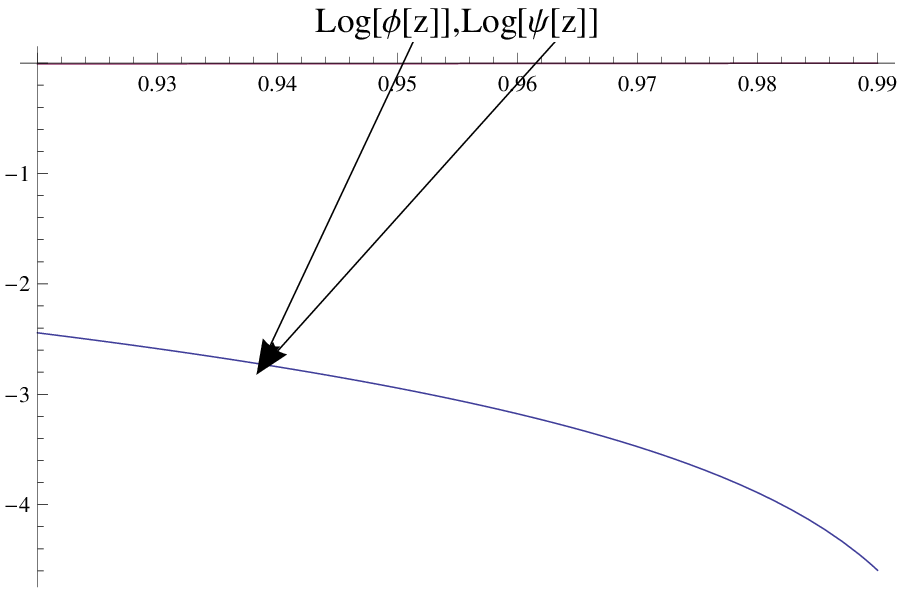} \\
\includegraphics[width=7.5cm]{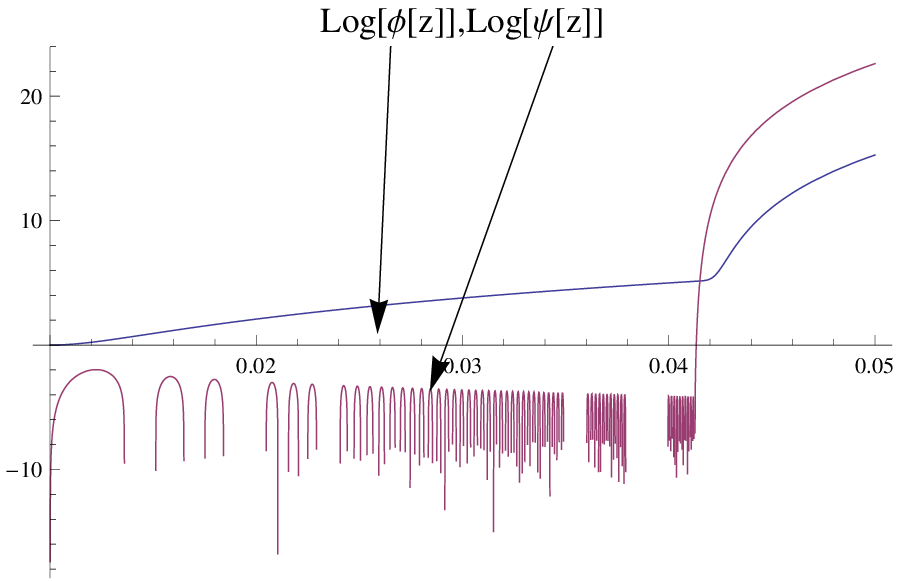}&
\includegraphics[width=7.5cm]{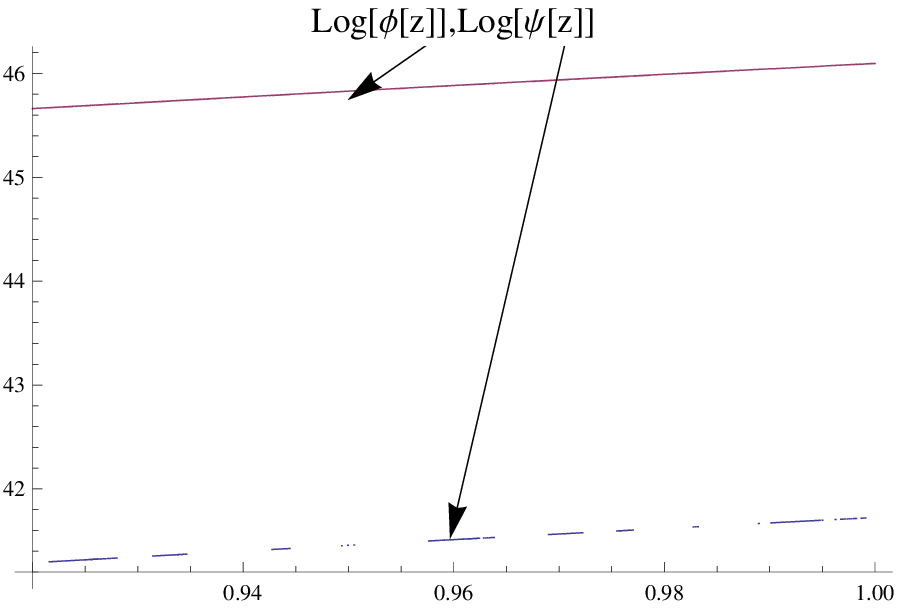} \\
\end{tabular}
\caption{ (\textit{ Left})  Field solutions near horizon boundary analytically.   (\textit{ Right}) Solutions near AdS boundary numerically for $m^2=-1$. }
\end{figure*}

\begin{figure*}[thbp]
\begin{tabular}{rl}
\includegraphics[width=7.5cm]{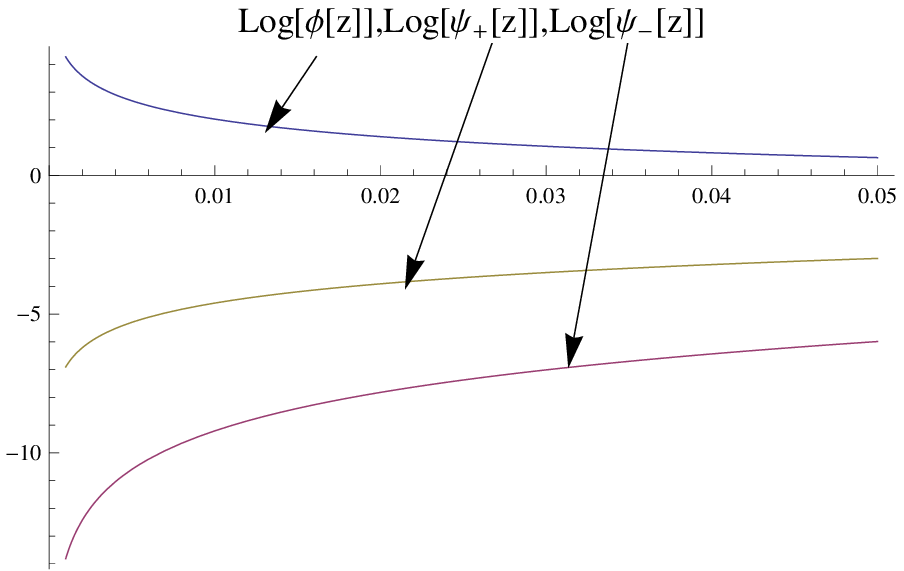}&
\includegraphics[width=7.5cm]{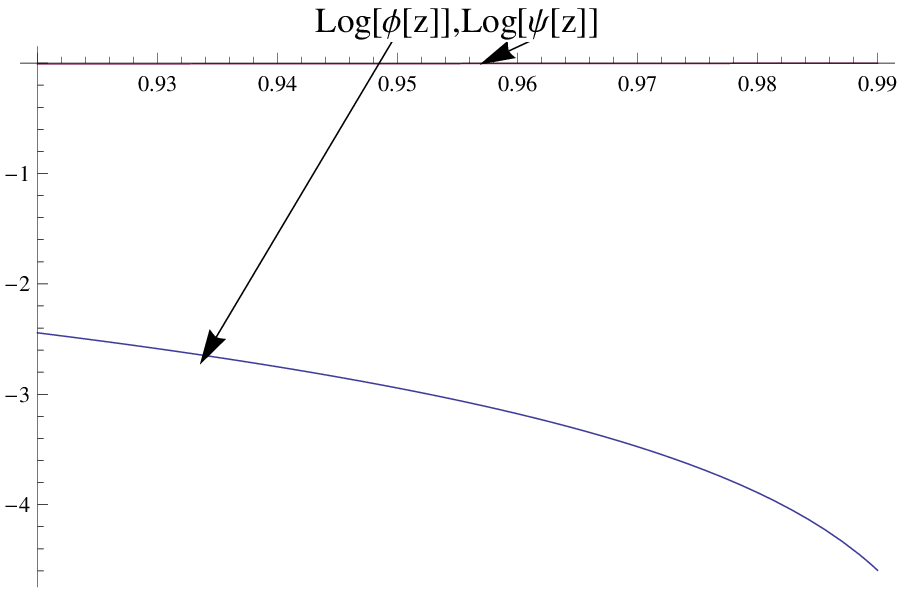} \\
\includegraphics[width=7.5cm]{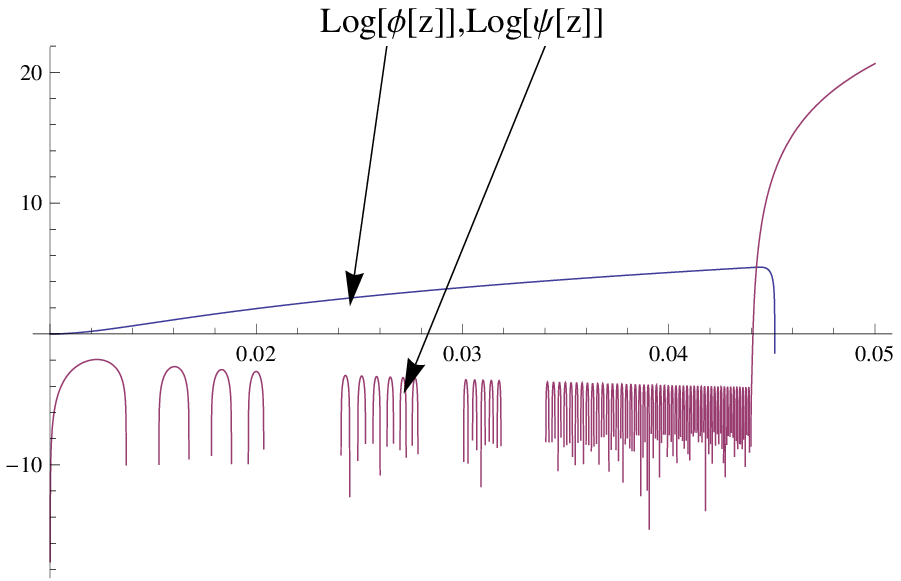}&
\includegraphics[width=7.5cm]{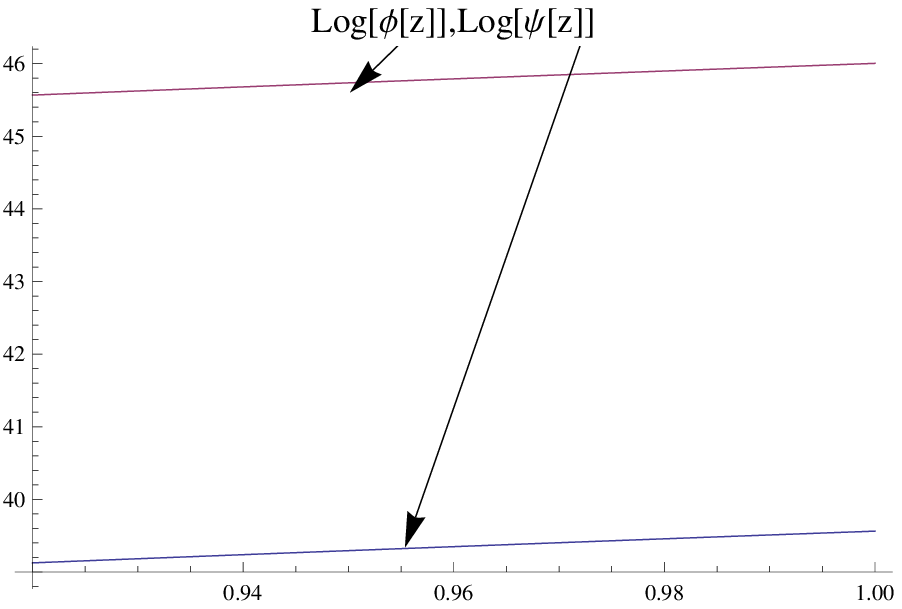} \\
\end{tabular}
\caption{ (\textit{ Left})  Field solutions near horizon boundary analytically.   (\textit{ Right}) Solutions near AdS boundary numerically for $m^2=-2$. }
\end{figure*}

\begin{figure*}[thbp]
\begin{tabular}{rl}
\includegraphics[width=7.5cm]{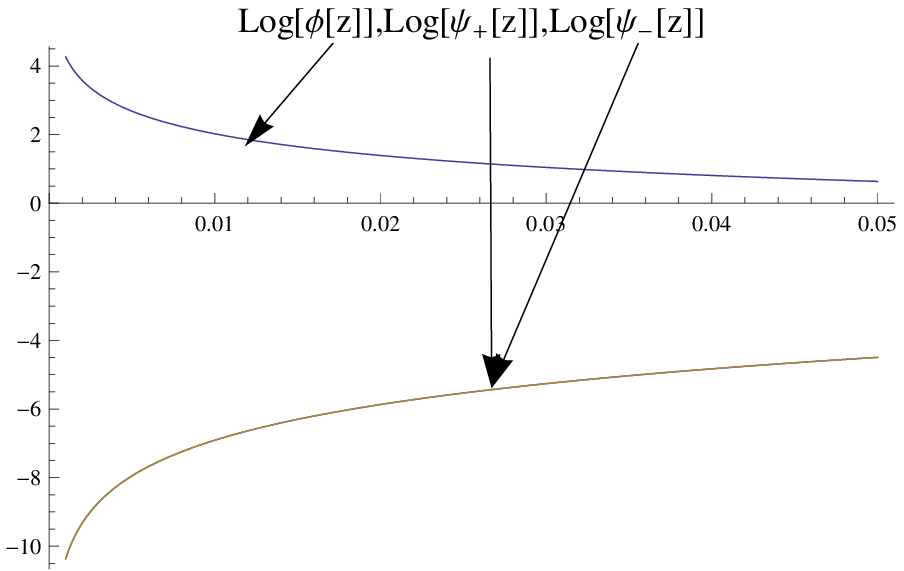}&
\includegraphics[width=7.5cm]{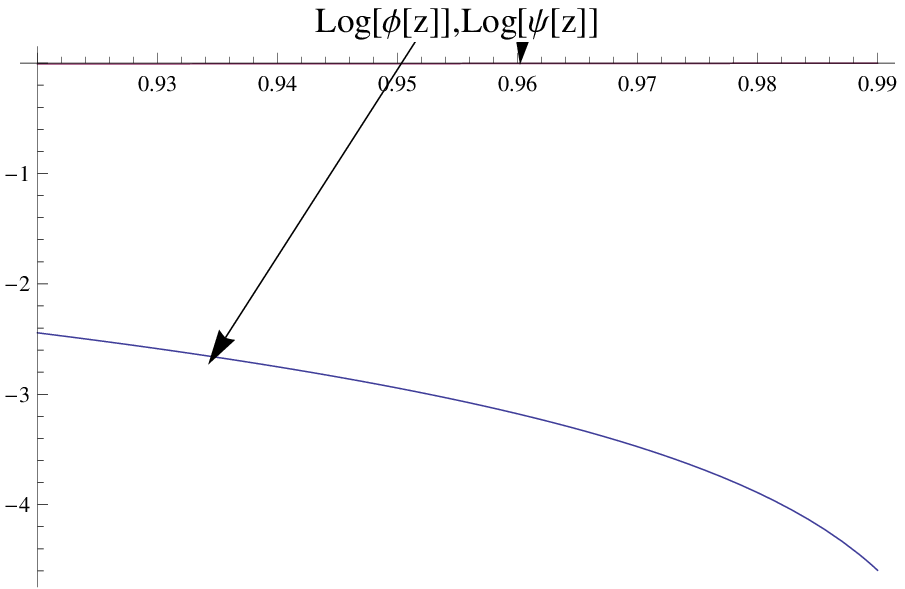} \\
\includegraphics[width=7.5cm]{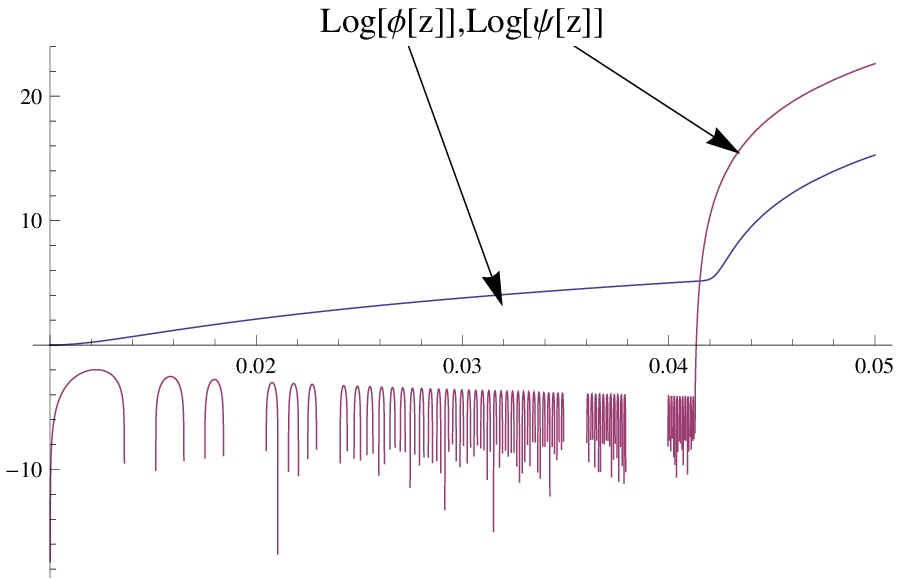}&
\includegraphics[width=7.5cm]{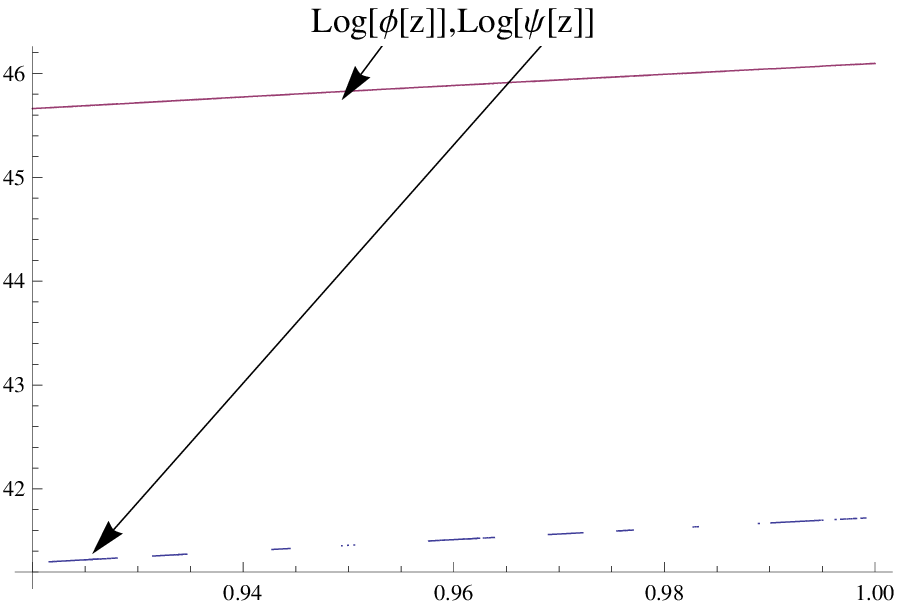} \\
\end{tabular}
\caption{ (\textit{ Left})  Field solutions near horizon boundary analytically.   (\textit{ Right}) Solutions near AdS boundary numerically for $m^2=-9/4$. }
\end{figure*}

\end{document}